\documentclass{ws-ijmpa}
\usepackage[super,compress]{cite}
\usepackage{graphicx}

\usepackage[koi8-r]{inputenc} 

\usepackage{float}
\usepackage{amssymb,amsmath}

\DeclareMathOperator{\arctanh}{arctanh}

\usepackage{epstopdf}
\usepackage{xcolor}

\begin{document}
\markboth{I. Bormotova, E. Kopteva, M. Churilova \&  Z. Stuchlik}{Accelerated expansion of the universe from the perspective of inhomogeneous cosmology}

\catchline{}{}{}{}{}
%
%%%%%%%%%%%%%%%%%%%%%%%%%%%%%%%%%%%%%%%%%%%%%%%%%%%%%%%%%%%%%%%%%%%%

\title{Accelerated expansion of the universe from the perspective of inhomogeneous cosmology}

\author{Irina Bormotova$^{1,2,^*}$, Elena Kopteva$^{1}$, Mariia Churilova$^{1}$ and Zdenek Stuchlik$^{1}$}
\address{$^{1}$Institute of Physics and Research Centre of Theoretical Physics and Astrophysics, Faculty of Philosophy and Science, Silesian University in Opava, Bezrucovo namesti 13, Opava, 746~01, Czech Republic\\
$^{2}$Bogoliubov Laboratory of Theoretical Physics, Joint Institute for Nuclear Research, Joliot-Curie 6, Dubna, 141980, Moscow Region, Russia\\
$^*$q\_Leex@mail.ru}

\maketitle

\begin{history}
\received{Day Month Year}
\revised{Day Month Year}
\end{history}

\begin{abstract}
We present a special case of the Stephani solution with spherical symmetry while considering different values of spatial curvature. We investigate the dynamics of the universe evolution in our model, build the R--T-regions for the resulting spacetime and analyze the behavior of the deceleration parameter. The singularities of the model are also discussed. The geometry of the spatial part of the obtained solution is explored.
\end{abstract}

\ccode{PACS numbers: 04.20.Jb, 04.20.-q}

\section{Introduction}
In recent years the cosmologists have considered Stephani solution \cite{Stephani67} as an alternative to the LambdaCDM model \cite{Stelmach08, Balcerzak15}. Such models could serve as a possible explanation of the accelerated expansion of the universe, besides the exotic forms of matter like dark energy. \cite{Krasinski83, Sussman00, Korkina16, Ong18}. 

In present work, we consider the Stephani solution in case of spherical symmetry. In Sec.~\ref{Sec2} we introduce this solution for our model and fit it to the current values of the cosmological parameters. The geometry of the spatial part and the dynamics of the universe evolution of the obtained solution are explored in Sec.~\ref{Sec3}. The Sec.~\ref{Sec4} is dedicated to conclusions.   

\section{The Model of the Universe with the Accelerated Expansion}\label{Sec2}
The Stephani solution is a non-static spherically-symmetric solution of the Einstein equations for the universe filled with shear-free perfect fluid with uniform energy density $\varepsilon=\varepsilon(t)$ and non-uniform pressure $p=p(\chi, t)$. It is well known that this solution contains the Friedmann solution as a particular case. Therefore, we require our model to give the flat Friedmann spacetime in the limit of zero curvature. In connection to this, we choose the energy density in the same form as for the Friedmann dust $\varepsilon =a_0/a^3$, $a_0=\mathrm{const}=a(t_0)$, where $a(t)$ is an analogue of the Friedmann scale factor; $t_0$ corresponds to the present time. All quantities are expressed in dimensionless units, where $c \equiv 1$ and $ 8\pi \gamma \equiv 1$. Thus, our model of the universe with the accelerated expansion in comoving coordinates is written
\begin{align}\label{ourm}
&\mathrm{d}s^2=\frac{\dot{r}^2 a^2}{r^2 \dot{a}^2}\mathrm{d}t^2-r^2\left(\mathrm{d}\chi^2+\chi^2\mathrm{d}\sigma^2\right),\\
& r=\frac{a}{1+\zeta a^2 \left(\frac{\chi}{2}\right)^2}, \qquad \zeta(t)=-|\beta| \frac{a_0^k} {a^{k+2}},
\end{align}
where $\mathrm{d}\sigma$ is the ordinary metric on a unit 2-sphere; dot represents the partial derivative with respect to time; $\zeta(t)$ is the spatial curvature; $k=\mathrm{const}; \beta=\mathrm{const}<0$, i.e., the spatial curvature is negative everywhere in the universe. 

The model contains three true singularities: the initial singularity ($r=0$) corresponding to the Big Bang, the singularity arising from the metric coefficient $g_{11}$: $\chi=\frac{2\left( \frac{a}{a_0} \right)^{\frac{k}{2}}}{\sqrt{|\beta|}}$ and the singularity originating from the pressure $p$: $\chi=\frac{2\left( \frac{a}{a_0} \right)^{\frac{k}{2}}}{\sqrt{(1+k)|\beta|}}$, which for the constant $k=-1$ brings us to the Stephani--Dabrowski model \cite{Dabrowski93} and for $k<-1$ no longer exhibits singular behavior.
\begin{table}[t]
\tbl{Cosmological parameters in terms of $\chi$ and $a(t)$.}
{\begin{tabular}{@{}ccc@{}} \toprule
\textbf{Hubble parameter} & \textbf{Matter density parameter} & \textbf{Universe radius}  \\ \colrule
$H=c\sqrt{\frac{a_0}{a^3}+|\beta|\frac{a_0^k}{a^{k+2}}}$ & $\Omega_{\mathrm{m}}=\frac{\varepsilon}{\varepsilon_{\mathrm{cr}}}=\frac{a_0 c^2}{3 a^3 H^2} $ & $r_0=\frac{4a_0}{|\beta|} \arctanh (\frac{\chi_0}{2}|\beta|)$\\ \botrule \\
\textbf{Pressure} & \multicolumn{2}{c}{\textbf{Deceleration parameter}}  \\  \colrule  $p=\frac{a_0}{a^3} \frac{\left(\frac{\chi}{2}\right)^2|\beta|k}{\left(\frac{a}{a_0}\right)^k-\left(\frac{\chi}{2}\right)^2|\beta|(k+1)}$  & \multicolumn{2}{c}{$q=\frac{k |\beta|\left( \frac{\chi}{2} \right)^2 }{\left( \frac{a}{a_0} \right)^k-(k+1)|\beta| \left( \frac{\chi}{2} \right)^2} \left[ 1 + \frac{ \left( \frac{a}{a_0} \right)^{k-1} +k|\beta| }{2\left(\left( \frac{a}{a_0} \right)^{k-1} + |\beta| \right)} \right]$} \\ \botrule
\end{tabular}} \label{table} 
\end{table}

In Table~\ref{table} we express some cosmological parameters in terms of $\chi$ and $a(t)$, which parametrize the time coordinate as $T \equiv \frac{a(t)}{a_0}$ and, as a result, we see that the accelerated expansion of the universe can occur due to the non-uniform pressure. In accordance with the observational data of WMAP \cite{WMAP} we obtained the numeric constants of the model related to the present time 

 $\qquad \qquad a_0=1.58\times 10^{26} \mathrm{m}, \qquad \beta = -0.111113, \qquad \chi_0=2.59906.$

\section{Geometry and Universe Evolution}\label{Sec3}
In this section we investigate the evolution of the universe in the obtained model.Let us consider an observer located at the center of symmetry $\chi=0$ who observes the expanding spheres representing the successive shells of the fluid labelled with the corresponding values of the coordinate $\chi$. The velocity of the expansion of each sphere may be found as follows: $v(t,\chi)=\frac{\mathrm{d}}{\mathrm{d}t}\int_0^{\chi}r(t,\chi)\mathrm{d}\chi.$
\begin{figure}[t]
\centering 
	\begin{minipage}{0.4\linewidth}
 		\includegraphics[width=\linewidth]{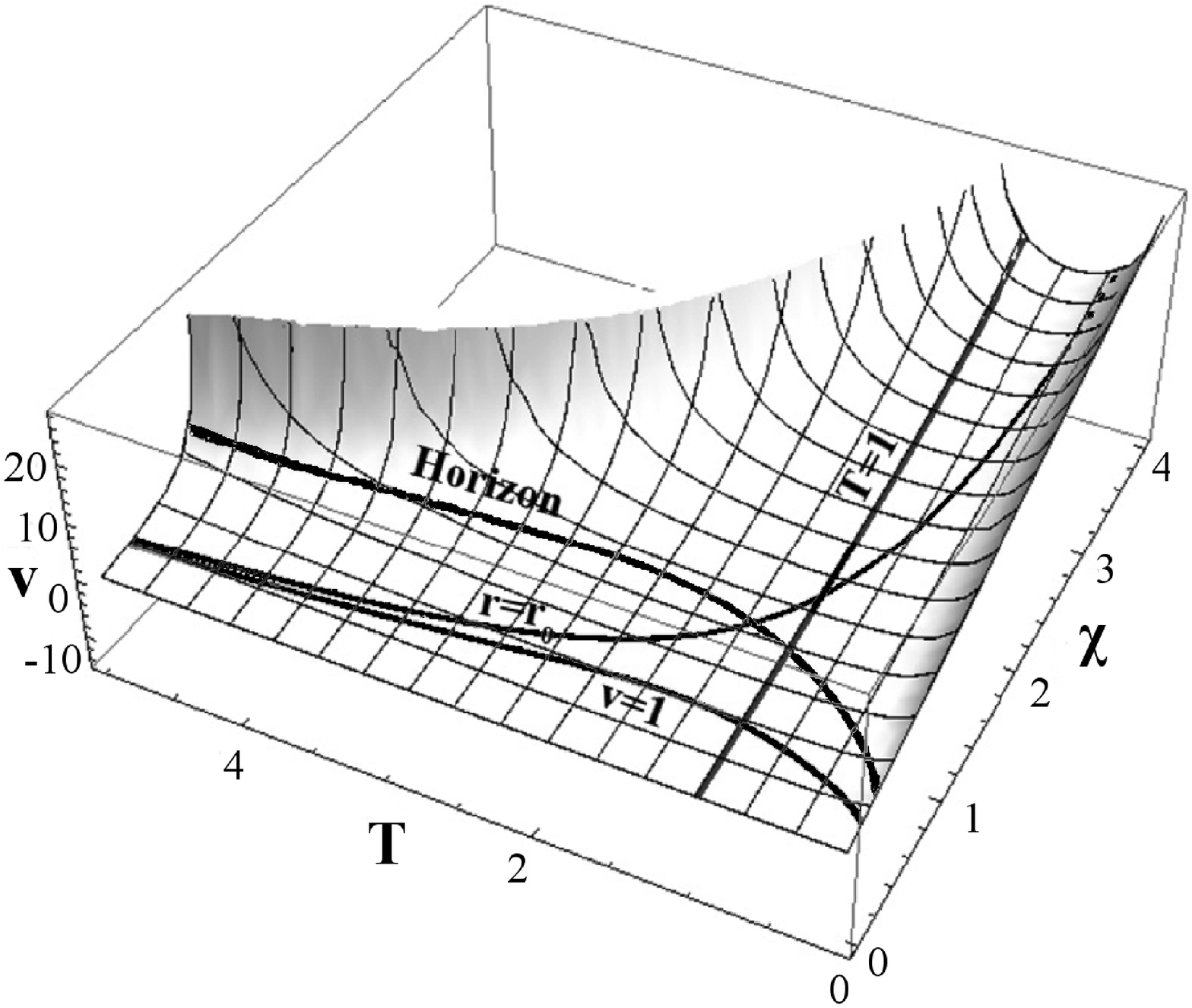}
	\end{minipage}
	\hspace{0.0005\linewidth}
 	\begin{minipage}{0.4\linewidth}
		\includegraphics[width=\linewidth]{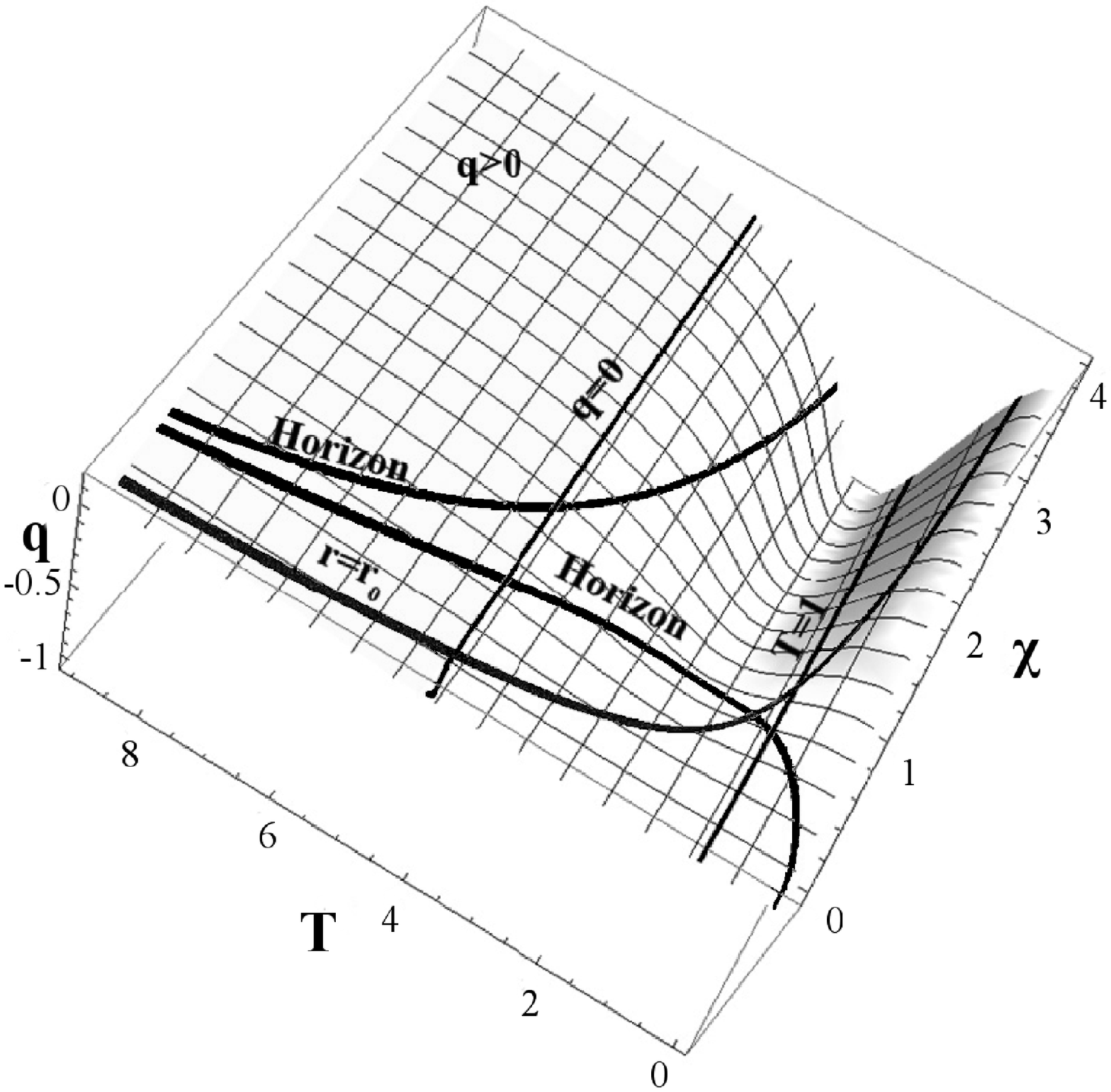}
	\end{minipage}
\caption{The profiles of the universe expansion velocity (left side) and deceleration parameter (right side) in the model for $k=-2.2$ in terms of dimensionless units, where the time parameter is defined as $T=a(t)/a_0$. This value of the index $k$ is chosen only for illustrative purposes, since with decreasing $k$ the picture qualitatively remains the same. The line $T=1$ corresponds to the present time. Left side: the Horizon line shows the horizon. T-region of essential non-stationarity of the spacetime is situated over the horizon. R-region corresponding to our world lies under the horizon. The central observer is located at the center of symmetry $\chi=0$ and always belongs to the R-region of the permitted observers. The line $r=r_0$ is the line along which the radius of the observable universe is equal to its current size found from the observations. The line $v=1$ indicates the so-called Hubble sphere (with radius $r(T,\chi))$ that expands with the speed of light. The point of intersection of the lines $T=1$ and $r=r_0$ defines the coordinate $\chi_0$ that indicates the sphere of radius $r_0$ corresponding to the edge of the universe. Right side: the Horizon lines are two branches of the horizon. T-region is situated between the branches of the horizon, while R-region is situated outside. For $k<-2$ there exists a line of zero deceleration parameter $q=0$. Hence one could expect that after some time the acceleration of the universe expansion changes into deceleration. However, even from the velocity profile (left side) it is clear that there is not any deceleration in the future. }
\label{fig1}
\end{figure}
\\
From Fig.~\ref{fig1} we can conclude that the universe at its present stage completely lies in the region which is not disturbed by any unphysical singularities.

\begin{figure}
\centering
	\begin{minipage}{0.5\linewidth}
		\includegraphics[width=\linewidth]{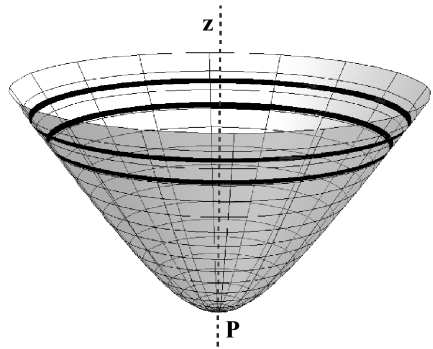}
	\end{minipage} 	
\caption{The spatial section $t=t_0$, $\theta=\pi/2$ of the spacetime (\ref{ourm}). The observer is situated at the point \textbf{P}.  The upper line is the line $r=r_0$ indicating the current size of the universe. The lower line is one branch of the horizon.}
\label{fig2}
\end{figure}

To investigate the spatial geometry of the obtained solution, we build the spatial sections of the spacetime with metric (\ref{ourm}), we fix the time at present moment $t=t_0=\mathrm{const}$ $(a=a_0=\mathrm{const})$. In order to reduce the number of spatial dimensions we also fix $\theta=\pi/2$ and embed the obtained hypersurface into the three-dimensional Euclidean space with cylindrical coordinates, which is shown in Fig.~\ref{fig2}.

To restore the three-dimensional picture one should imagine that the circles along the sections with $z=\mathrm{const}$ are in fact 2-spheres. It turns out that the form of spatial sections taken at present time does not depend on the  power of $a(t)$ in our chosen curvature function $\zeta(t)$. 

\section{Conclusions}\label{Sec4}
We analyzed various properties of the spherically-symmetric Stephani model, investigated its spatial geometry, and compared its predictions with the available observational data. The R--T-structure of the obtained spacetime revealed that the central observer will always belong to the R-region of the permitted observers. In the future the accelerated expansion of the universe will not change into deceleration.

The obtained results provide evidence that such solution could successfully model the current stage of evolution of the universe without any harm from existing singularities and without introducing any exotic types of matter.

\section*{Acknowledgments}
This work supported by the Grant of the Plenipotentiary Representative of the Czech Republic in JINR under Contract No.~208 from 02/04/2019. The authors acknowledge the Research Centre of Theoretical Physics and Astrophysics of the Faculty of Philosophy and Science, Silesian University in Opava for support. Z.~S.\ acknowledges the Albert Einstein Centre for Gravitation and Astrophysics supported by the Czech Science Foundation Grant No.~14-37086G. I.~B.\ acknowledges the Silesian University in Opava grant SGS 12/2019.

\end{document}